\documentclass[journal]{vgtc}       

\usepackage{enumitem}
\usepackage{paralist}
\usepackage{balance} 
\usepackage{color}
\usepackage[dvipsnames,svgnames]{xcolor}
\usepackage[most]{tcolorbox}
\onlineid{1014}

\newcommand\eg{e.g.}

\newcommand{\clusterview}{\textit{Cluster View}\xspace}
\newcommand{\projectionview}{\textit{Projection View}\xspace}
\newcommand{\paramview}{\textit{Variable View}\xspace}
\newcommand{\subgroupview}{\textit{Social Group View}\xspace}
\newcommand{\lineup}{\textit{LineUp Table}\xspace}
\newcommand{\mapview}{\textit{Map View}\xspace}
\newcommand{\detailview}{\textit{Detail View}\xspace}
\newcommand{\corrmatrix}{\textit{Correlation Matrix}\xspace}
\newcommand{\pcp}{\textit{Parallel Sets Chart}\xspace}
\newcommand{\projbar}{\textit{Projection Bar}\xspace}
\newcommand{\projbars}{\textit{Projection Bars}\xspace}
\newcommand{\streetview}{\textit{Street View}\xspace}
\newcommand{\systemname}{\textit{NeighViz}\xspace}

\definecolor{secondround}{RGB}{0,0,0} 
\newcommand{\secondround}[1]{\textcolor{secondround}{#1}}

\usepackage{academicons}
\definecolor{orcidlogocol}{HTML}{A6CE39}

\vgtccategory{Research}

\title{\systemname: Towards Better Understanding of Neighborhood Effects on Social Groups with Spatial Data}

\author{%
    Yue Yu, 
    Yifang Wang*, 
    Qisen Yang, 
    Di Weng, 
    Yongjun Zhang, 
    Xiaogang Wu, 
    Yingcai Wu, 
    and Huamin Qu
}

\authorfooter{
  \item Yue Yu and Huamin Qu are with The Hong Kong University of Science and Technology.
  	E-mail: yue.yu@connect.ust.hk, huamin@cse.ust.hk. 
  \item Yifang Wang is with The Center for Science of Science and Innovation, Northwestern University. 
        E-mail: yifang.wang@kellogg.northwestern.edu. 
  \item Qisen Yang and Yingcai Wu are with the State Key Lab of CAD\&CG, Zhejiang University. 
        E-mail: \{qs\_yang, ycwu\}@zju.edu.cn.
  \item Di Weng is with the School of Software Technology, Zhejiang University. 
        E-mail: dweng@zju.edu.cn.
  \item Yongjun Zhang is with Stonybrook University. 
        Email: yongjun.zhang@stonybrook.edu. 
  \item Xiaogang Wu is with New York University. 
        Email: xw29@nyu.edu. 
  \item Yifang Wang is the corresponding author.
}

\abstract{
Understanding how local environments influence individual behaviors, such as voting patterns or suicidal tendencies, is crucial in social science to reveal and reduce spatial disparities and promote social well-being. 
With the increasing availability of large-scale individual-level census data, new analytical opportunities arise for social scientists to explore human behaviors (\eg, political engagement) among social groups at a fine-grained level.
However, traditional statistical methods mostly focus on global, aggregated spatial correlations, which are limited to understanding and comparing the impact of local environments (\eg, neighborhoods) on human behaviors among social groups. 
In this study, we introduce a new analytical framework for analyzing multi-variate neighborhood effects between social groups.
We then propose \systemname, an interactive visual analytics system that helps social scientists explore, understand, and verify the influence of neighborhood effects on human behaviors.
Finally, we use a case study to illustrate the effectiveness and usability of our system.
} 


\graphicspath{{figs/}{figures/}{pictures/}{images/}{./}} 

\usepackage{tabu}                      
\usepackage{booktabs}                  
\usepackage{lipsum}                    
\usepackage{mwe}                       
\usepackage{setspace}

\usepackage{mathptmx}                  

\begin{document}
\maketitle
\begin{spacing}{1} 
\section{Introduction}
Spatial data is prevalent in various social science disciplines, such as political science, sociology, and public health. 
The spatial differences in the correlations among variables (\eg, demographic and socioeconomic variables) have raised numerous research questions, particularly in the studies of neighborhood effects~\cite{publicHousing} and social group comparisons~\cite{negativitygap_2010, rajcolleges}. 
For example, poor Americans exposed to neighbors from a broader range of socioeconomic classes tend to have better financial outcomes ~\cite{socialcaptial1, socialcaptial2}. 
Likewise, neighborhood centers promoting social interaction among the elderly are associated with reducing depressive symptoms, especially in low socioeconomic neighborhoods ~\cite{Miao2022Promoting}.

Traditional neighborhood effect analysis in social science is primarily hypothesis-driven with a focus on a broad social group at a coarse-grained level (\eg, the elderly in one particular city) due to limited high-granular datasets ~\cite{Wu-suicide}. 
However, the recent availability of large-scale individual-level geospatial datasets (\eg, L2 Voter and Consumer Data \cite{l2}) has provided experts with new opportunities for analyzing detailed neighborhood effects across social groups, such as partisan segregation in activity space ~\cite{zhang2023human} and the adoption of prosocial behavior in different partisan areas ~\cite{baxter2022local}. 
Nevertheless, the expansion in data volume and the high diversity of variables introduce new data-driven analytical demands for variable selection, spatial modeling, and comparative analysis between groups. 
Social scientists who adopt such datasets often face challenges in exploring, interpreting, and comparing the modeled neighborhood effects among diverse social groups in an effective approach. 
\secondround{For instance, scholars may observe a negative association between neighborhood socioeconomic status and voting participation at the aggregated level in the model results, but they cannot easily understand the variations across neighborhoods and social groups from numbers.
Better tools are desired to help social scientists dive into specific contexts and examine how voting participation differs across neighborhoods.}

Visual analytics thus offers a promising solution to overcome the aforementioned analytical demands by utilizing intuitive visual representations and interactions.
However, developing a visualization system for analyzing neighborhood effects over various social groups still poses three challenges.
First, conventional social science methodologies lack a coherent workflow for multivariate spatial analysis that effectively surfaces neighborhood effects on social groups. 
Most approaches involve multiple separate models and tools (\eg, statistical and geographic information systems (GIS) software), which require much effort to go back and forth between different analytical steps.  
Second, visually presenting the complex spatial and social relationships among different neighborhoods and social groups is difficult. 
Previous work has focused on visualizing either spatial patterns~\cite{garcia2019crimanalyzer, weng2020towards} or multivariate social groups~\cite{wang2021seek}. 
It is essential to bridge the gap between these approaches with unified representations that support the effective analysis of both spatial and multivariate social group data.
Third, designing a visualization system to support social scientists in exploring, analyzing, and verifying insights in an interactive and rigorous manner is a non-trivial task. 
Multiple coordinated views are required to support data-driven and intuitive exploration. 
Moreover, providing the experts with contextual details is also necessary to verify their findings. 

To address the first challenge, we formulate a data abstraction (Section \ref{data_desc}) and characterize the problem domain of neighborhood effects on social groups with our experts. 
We then propose an analytical framework combining data-driven techniques with domain-specific models.
For the second challenge, we apply visualization techniques (\eg, 1D Map Projection and Parallel Sets) to reveal neighborhood effects and inter-group differences, enhancing comprehension of complex spatial and social relationships.  
For the third challenge, we present \systemname, a visual analysis system that aids social scientists in modeling, exploring, and verifying neighborhood effects on social groups across social science issues. 
We evaluate \systemname through a case study with a domain expert to showcase its effectiveness and usability. 
\begin{figure*} [t]
 \centering 
 \vspace{-0.5cm}
 \includegraphics[width=0.95\linewidth]{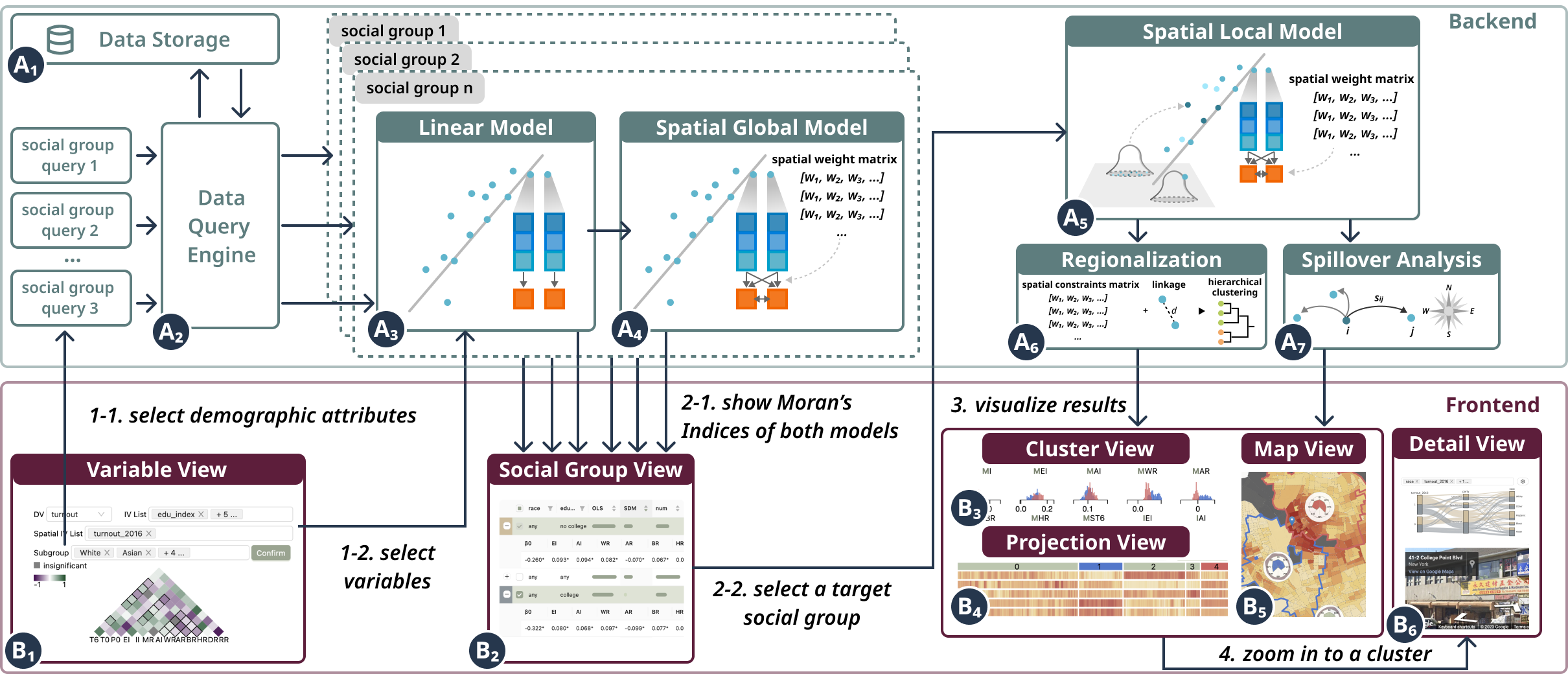} 
 \vspace{-0.25cm}
 \caption{
    \systemname comprises the backend (A1-A7) and the frontend modules (B1-B6). Data is initially preprocessed and stored in the database (A1) and then aggregated via a Data Query Engine (A2) based on user-selected demographic attributes (1-1) for social group analysis. The filtered data is then sent to the data analysis pipeline (A3-A7, Section~\ref{sec:03_analysis_framework}). The six visualization views (B1-B6, Section~\ref{sec:04_visual_design}) interact with backend components \secondround{through HTTP requests}, facilitating interactive data analysis \secondround{(1-1 to 4)}. 
 }
 \vspace{-0.6cm}
 \label{fig:framework}
\end{figure*}

\section{Requirement Analysis} 
\label{sec:02_Background_Requirement_Analysis} 
Incorporating both neighborhood effects and social group analysis is essential for a holistic understanding of social dynamics and the complex relationships between people and their environments.
To facilitate the generation of hypotheses about the neighborhood effect on social groups with the power of a visual analytic system, we have collaborated with two social scientists ($E_A$ and $E_B$) in sociology over the past year. 
Though our system primarily targets social scientists who lack proficiency in geospatial analysis, we also sought consultation with a GIS expert ($E_C$) about spatial modeling.
Based on their traditional analytical approaches, we have summarized a three-stage workflow for the modeling, exploration, and verification of the neighborhood effect on social groups as follows. 

\vspace{1mm}
\noindent \textbf{Model Generation.} 
In the first stage, experts aim to get an overview of potential factors for the specific social problem and identify a target social group for in-depth analysis: 

\noindent \textbf{T1: Selecting variables for the spatial model.} 
Experts usually start from the variable selection process with a global model (\eg, Ordinary Least Square (OLS)). 
The system should support both univariate and multivariate exploration via intuitive visualization. 
Additionally, it should also help detect issues lying in model robustness, such as multicollinearity and spatial autocorrelation. 

\noindent \textbf{T2: Identifying social groups for further exploration.}
Selecting a social group of interest from various possibilities (\eg, based on multiple demographic attributes, such as Whites with college education) based solely on domain knowledge can be challenging. 
A data-driven approach is thus essential to help experts identify groups based on the spatial heterogeneity of the spatial global model, and focus on it in the subsequent local-level analysis. 

\vspace{2mm}
\noindent \textbf{Geographical Exploration.}
After locating factors and social groups of interest, the experts will conduct local spatial analysis to identify and understand neighborhood effects: 

\noindent \textbf{T3: Exploring geographical distribution.} 
Scalable visualization should be designed to reveal the various spatial attributes (\eg, multi-dimensional raw data and outputs of the spatial model) and facilitate comparisons through geo-distribution. 
Conventional 2D maps are difficult to compare multiple variables intuitively. 

\noindent \textbf{T4: Detecting areas with potential neighborhood effect.} 
One core task for experts is to detect and explore neighborhood effects. 
Thus, the system should cluster spatial units into neighborhoods with shared characteristics and internal cohesion based on expert-focused attributes.
Moreover, novel visualization is required to reveal contextual information about each neighborhood, such as the statistical summary and spatial spillover \cite{sampson2012great}, to help understand and interpret such effects. 

\vspace{2mm} 
\noindent \textbf{Comparison and Verification.} 
Based on spatial patterns, the next step is to drill down to additional contexts beyond the spatial model and statistics to interpret specific neighborhood effects. 
However, experts usually need to switch between multiple data sources and software to gain such insights, which is laborious and inefficient: 

\noindent \textbf{T5: Comparing across social groups.} 
To comprehend the driving factors of the neighborhood effects on a specific social group, experts want the tools to support the comparison of different social groups. 
The system should use visual comparison techniques to help compare groups with multiple attributes. 

\noindent \textbf{T6: Verifying and explaining neighborhood effects.}  
In addition to group comparison, qualitative geo-information is also important for the expert to learn the neighborhood environments and interpret the factors affecting the residents' behaviors. 
\section{Neighborhood Effect Analysis Framework}
\label{sec:03_analysis_framework} 

\begin{figure*}[t]
 \centering 
 \vspace{-0.5cm}
 \includegraphics[width=0.95\linewidth]{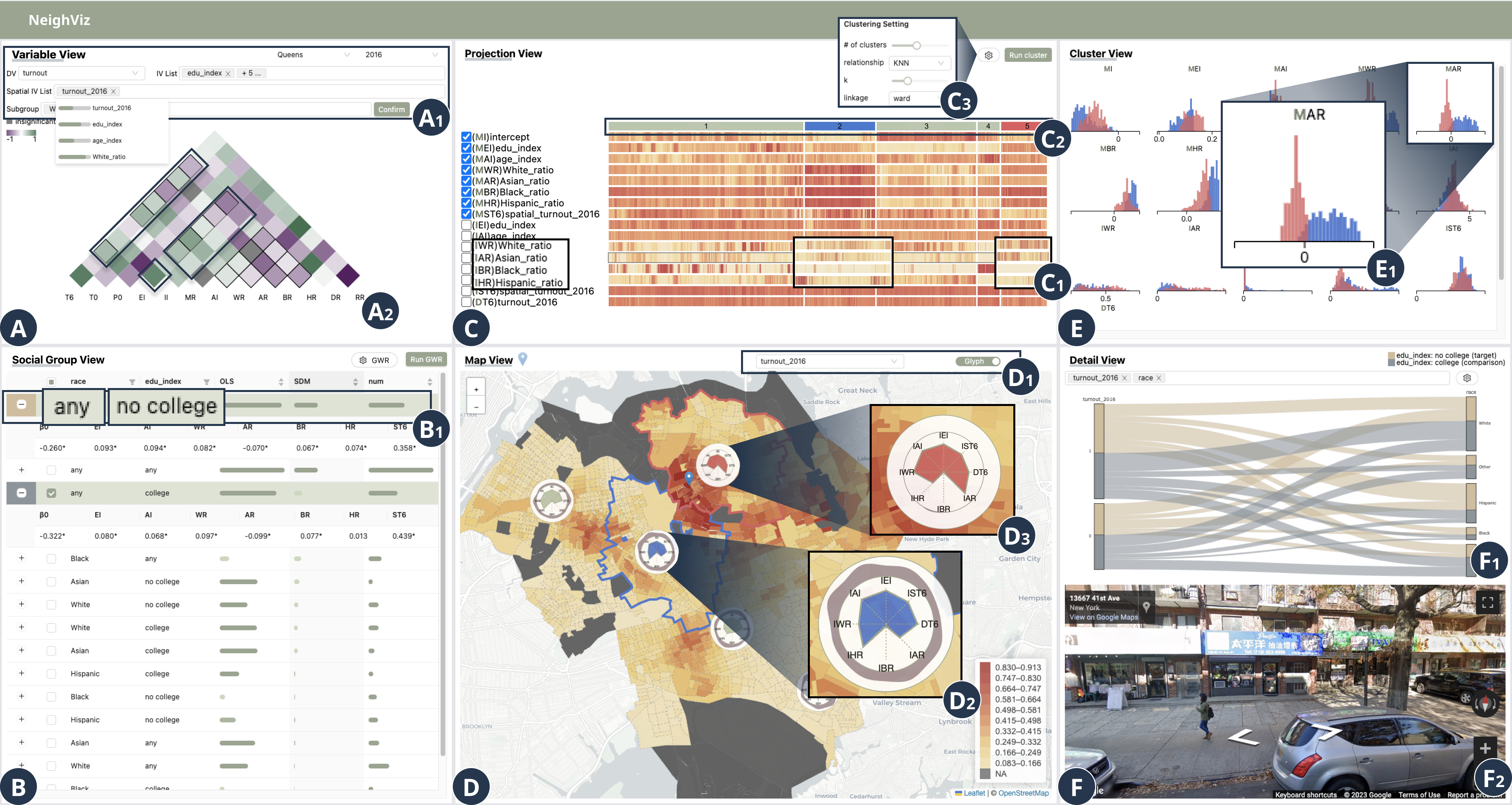} 
 \vspace{-0.2cm}
 \caption{
     The system interface of \systemname. 
     (A) The \paramview supports the variable selection for the spatial model using the \corrmatrix.
     (B) The \subgroupview summarizes global model results and Moran's Indices for social groups to help group selection.
     (C) The \projectionview offers a regionalized overview of spatial distributions of attribute values and spatial local model coefficients in 1D \projbars.
     (D) The \mapview uses a choropleth map to show the spatial distribution of a selected variable, with cluster glyphs summarizing variable statistics and spillover effects of neighborhood clusters. 
     (E) The \clusterview lists variable distributions of different neighborhood clusters. 
     (F) The \detailview provides detailed information of a specific cluster via the \pcp and the street view.
 }
 \vspace{-0.6cm}
 \label{fig:visual disign-1}
\end{figure*}

We propose a web-based application consisting of a backend and a frontend (see Fig.~\ref{fig:framework}) for analyzing neighborhood effects on social groups \secondround{(more implementation details available in Section \ref{implementation})}. 
\secondround{This section focuses on the data analysis pipeline in the backend (Fig.\ref{fig:framework}-A3 to A7), which combines conventional and advanced social science analysis methods. We developed the pipeline based on the social science literature~\cite{wen2018evaluating, bulteau2018spatial, zhu2020impacts, zhao2022integrated} with our domain experts.}

\label{sec:04_Data_Analysis_Preprocassing} 

\textbf{Linear Model.}
We initially use ordinary least-square (OLS) as the baseline model to explore variable relationships (Fig.~\ref{fig:framework}-A3), where $y$ represents the dependent variable and $X$ represents the independent variables selected by users (Fig.~\ref{fig:framework}-1-2).
\begin{equation}
  y = X\beta + \epsilon
\end{equation}
However, OLS treats spatial units as independent observations, neglecting spatial dependencies by neighborhood effects.
To measure spatial dependencies in the model, we compute Moran's Index ~\cite{moran1950notes}. A high Moran's Index indicates unaccounted spatial dependencies. 

\textbf{Spatial Global Model.}
To account for spatial dependencies, we transition from the OLS to Spatial Durbin Model (SDM) \cite{lesage2009introduction}. The SDM equation includes additional terms for the spatial lag in dependent ($\rho Wy$) and independent variables ($\gamma W X\beta$), reflecting the influence from neighboring spatial units. A spatial weight matrix ($W$) determines the extent of this influence, \secondround{which is the Gaussian kernel by default but also supports various kernel-based and contiguity-based weighting methods}.

\begin{equation}
y = X\beta + \rho Wy + \gamma W X\beta + \epsilon
\end{equation}
However, as a global model, SDM cannot effectively address spatial heterogeneity due to location-specific factors (\eg, culture and history).
Consequently, Moran's Index is utilized again to detect spatial heterogeneity in SDM results. 
The Moran's Indices for both OLS and SDM (Fig.\ref{fig:framework}-2-1) of all the social groups will be computed and displayed in the frontend (Fig.\ref{fig:framework}-B2). 
Users can rank groups based on these indices and choose a group of interest for further local analysis (Fig.\ref{fig:framework}-2-2). 

\textbf{Spatial Local Model.} 
To explore spatially varying neighborhood effects over places, we extend the SDM to the Geographically Weighted Regression (GWR) model ~\cite{brunsdon1998geographically} (Fig.~\ref{fig:framework}-A5).
GWR, a local regression technique, estimates a set of regression coefficients for each spatial unit, considering spatial heterogeneity and Tobler's First Law of Geography ~\cite{tobler1970computer} with the following equation. 

\begin{equation}
y_i = X_i\beta_i(u_i, v_i) + \rho W_i y_i + \gamma_i W_i X_i\beta_i(u_i, v_i) + \epsilon_i
\end{equation}

The GWR model is similar to OLS and SDM while including \secondround{$\beta_i(u_i, v_i)$, a function that depends on the spatial coordinates of spatial unit $i$ ($u_i$ and $v_i$). It estimates the spatially varying coefficients at each location by giving more weight to nearby observations and less weight to more distant observations.}
The GWR model generates unique coefficients per variable per spatial unit. These coefficients are used in the subsequent steps of Regionalization and Spillover Effect Analysis.

\textbf{Regionalization.}
To assist in the identification of areas that demonstrate potential neighborhood effects, we apply regionalization to process outputs of the spatial local model, where each spatial unit is characterized by a model coefficient for each variable.
Regionalization groups spatial units into regions based on similar attribute values and model coefficients, thereby facilitating the discovery of latent neighborhoods with similar characteristics (Fig.~\ref{fig:framework}-A6).
We implement spatially constrained hierarchical clustering, which combines elements of hierarchical clustering with spatial constraints, ensuring that the resulting clusters are both internally homogeneous and spatially contiguous. 
The number of clusters is set to $5$ by default and can be adjusted by users. 

\textbf{Spillover Effect Analysis.} 
\label{spillover}
To explore the dynamics of the spillover effect in different regions, a local spillover effect algorithm has been developed.
This algorithm utilizes the coefficients of the spatially lagged variables in the spatial local model as its input.
It computes the magnitude and direction of the spillover effect from each spatial unit to its neighboring units. (Fig.~\ref{fig:framework}-A7). 
For each neighboring unit $j$ of the focal unit $i$, we multiply the coefficient of spatially lagged variables, $\gamma_{j}(u_j, v_j)$, with the weight $W_{ij}$:

\begin{equation}
S_{ij} = \gamma_{j}(u_j, v_j) \cdot W_{ij}
\end{equation}
We then calculate the relative direction of $j$ to $i$ and categorize the direction into one of the 16 cardinal directions. We finally aggregate the strength of the spillover effect of variables in 16 cardinal directions for the focal spatial unit. 

All the analysis results in the pipeline are fed into the six visualization views (Fig.~\ref{fig:framework}-B1 to B6) for interactive analysis.

\section{Visual Design}
\label{sec:04_visual_design}

\systemname features six views to support the analytical tasks in Section \ref{sec:02_Background_Requirement_Analysis}.
The expert can start from the \paramview (Fig. \ref{fig:visual disign-1}-A) to select variables and generate spatial models based on the multivariate analysis using the \corrmatrix (Fig. \ref{fig:visual disign-1}-A2) (\textbf{T1}). 
Next, the expert can compare the model results of different social groups in the \lineup \cite{gratzl2013lineup} in the \subgroupview (Fig. \ref{fig:visual disign-1}-B), and select an interesting group for further analysis (\textbf{T2}).
Through \projectionview (Fig. \ref{fig:visual disign-1}-C) and \mapview (Fig. \ref{fig:visual disign-1}-D), the expert can explore the geographical information on the clusters that are regionalized based on the model's coefficients (\textbf{T3, T4}).
Meanwhile, the \clusterview (Fig. \ref{fig:visual disign-1}-E) is provided to investigate the detailed distributions of variables with multiple density histogram charts. 
Finally, the expert can zoom into a cluster of interest and utilize the \detailview (Fig. \ref{fig:visual disign-1}-F) for in-depth details. \detailview supports social group comparisons through randomly sampled individual data displayed on \pcp \cite{inselberg1990parallel} (Fig. \ref{fig:visual disign-1}-F1), and neighborhood context exploration via Google Street View \cite{streetview} (Fig. \ref{fig:visual disign-1}-F2) (\textbf{T5, T6}).
In the following section, we introduce two main views, \projectionview and \mapview, in detail. 

\subsection{Projection View}
The \projectionview (Fig. \ref{fig:visual disign-1}-C) uses 1D \projbars to show the spatial distribution of attributes in regionalized clusters \textbf{(T3)}.

\textit{Description:} each \projbar (Fig. \ref{fig:visual disign-1}-C1) corresponds to a variable and is partitioned based on regionalization, with cluster segment bars showing the scope of clusters at the top (Fig. \ref{fig:visual disign-1}-C2). 
We apply binary tree traversal to generate \projbars as a dimensionality reduction approach~\cite{franke2021visual}.
Specifically, we use agglomerative hierarchical clustering for the regionalization (Section~\ref{sec:03_analysis_framework}). 
Then, we use the leaf order of the dendrogram to project the variables from the 2D map to 1D \projbars.
The normalized value of each spatial unit is encoded using a color scale from light yellow (lowest) to dark red (highest). 
Users can customize regionalization parameters to fine-tune the results (Fig. \ref{fig:visual disign-1}-C3). 
They can also click a \projbar to show the detailed distribution of a variable in the \mapview with the same color scheme.
\secondround{We first tried small-multiple maps to show variable distributions~\cite{wang2018visual, Hulstein2023Geo}. However, they became unscalable and hard to compare as the number of variables increased. Thus, we chose the projection method to show multivariate spatial distributions in a compact way.} 


\subsection{Map View}
Although the \projectionview shows variables compactly, it loses part of the spatial relationships. 
The \mapview thus provides more details with a 2D map (Fig. \ref{fig:visual disign-1}-D) \textbf{(T3)}, which includes cluster glyphs showing aggregated statistics and spillover effect in clusters \textbf{(T4)}. 

\textit{Description:} our design uses a choropleth map to show the spatial distribution of a chosen variable, with the same color encoding in the \projectionview.
We specifically designed a cluster glyph (Fig. \ref{fig:visual disign-1}-D2, D3) for each spatial cluster to show the statistics and illustrate the within-cluster spillover effect. 
The inner layer is a radar chart showing the mean values of selected variables across all spatial units of a cluster, with each axis representing the normalized mean of a variable. 
The outer layer reveals the spillover effect of the spatially lagged variables along $16$ cardinal directions as described in Section \ref{sec:03_analysis_framework}, \secondround{presented as a closed cardinal curve. The radius of the curve along each direction represents the mean magnitude of the spillover effect of the spatial units within that cluster,} and a larger radius indicates a stronger influence of the cluster along that direction. Users can click on a glyph to highlight other information about this cluster in multiple views (Fig.\ref{fig:framework}-4). 

\section{Case Study}
We applied \systemname to study relations between race and political engagement in the US. 
Specifically, we used data from the L2 Voter and Consumer Data \cite{l2} that contains 180 million registered voters in the US. 
We focused on New York City as a demonstration. 
To study neighborhood influence on political engagement, we used the voter turnout rate ($\frac{Number~of~votes~cast}{Total~number~of~eligible~voters}$) in general elections as the dependent variable. 
The data were aggregated to the census block group (CBG) level as described in Section~\ref{data_desc}.
We invited the political scientist $E_A$ (Section ~\ref{sec:02_Background_Requirement_Analysis}) to freely explore the system and finally organized his observations into a case as follows. 


\textbf{Model Generation.} 
$E_A$ began his study in Queens County for its cultural diversity, starting with the 2016 election data. Intrigued by the potential impact of racial segregation on the turnout rate, he selected the race ratios as independent variables, together with other potential factors, including education, income, and age. 
Then he noticed a high correlation between the \textit{Education Index (EI)} and \textit{Income Index (II)} in the \corrmatrix (Fig. \ref{fig:visual disign-1}-A2). To prevent multicollinearity, he excluded \textit{II}. He added a spatially lagged variable of the \textit{turnout rate} to study the voting behavior spillover effect. For social groups, he selected \textit{edu} and \textit{race} attributes to examine social group voting patterns. The final model and the resulting social group analysis are presented in the \subgroupview (\textbf{T1}).
The \lineup in the \subgroupview shows lower Moran's Indices for the Spatial Durbin Model (SDM) than the Ordinary Least Square (OLS), indicating better spatial dependence captured by the SDM. 
Noticing that the ``no college'' group had the highest Moran's I for SDM (Fig. \ref{fig:visual disign-1}-B1), $E_A$ inferred its spatial heterogeneity wasn't fully captured. Consequently, he employed the Geographically Weighted Regression (GWR) on this group to delve deeper into its local-level neighborhood effect. (\textbf{T2}).

\textbf{Geographical Exploration.}
The GWR model coefficients were then automatically regionalized into $5$ clusters, showing distinct correlations between independent and dependent variables. 
$E_A$ used glyphs in the \mapview (Fig. \ref{fig:visual disign-1}-D2, D3) to compare cluster statistics and quickly found that clusters 1 (blue) and 4 (red) were intriguing.
\textit{``Two adjacent regions: cluster 5 has an extremely high Asian Ratio, but cluster 2 looks more diverse. 
And why the spillover effect of cluster 2 is much stronger than that of cluster 5?''} 
He then clicked the two glyphs to highlight the clusters in \projectionview, \mapview, and \clusterview (\textbf{T3, T4}) to for more details. 
%
The \projectionview (Fig. \ref{fig:visual disign-1}-C1) confirmed the racial diversity in cluster 2 and racial segregation in cluster 5. 
Several model coefficients varied between clusters, leading him to conduct a closer comparison using \clusterview. 
The density histograms (Fig. \ref{fig:visual disign-1}-E1) showed a negative correlation between Asian concentration and less-educated voter turnout in cluster 5, while cluster 2 exhibited the reverse. $E_A$ commented it might illustrate the dynamics of the ethnic enclave theory \cite{hajnal2005turnout} and that community diversity could enhance political participation among less-educated minority populations, accounting for stronger voting spillover in cluster 2 (\textbf{T3, T4}). 

\begin{figure} [!htb]
 \centering 
 \vspace{-0.3cm}
 \includegraphics[width=\linewidth]{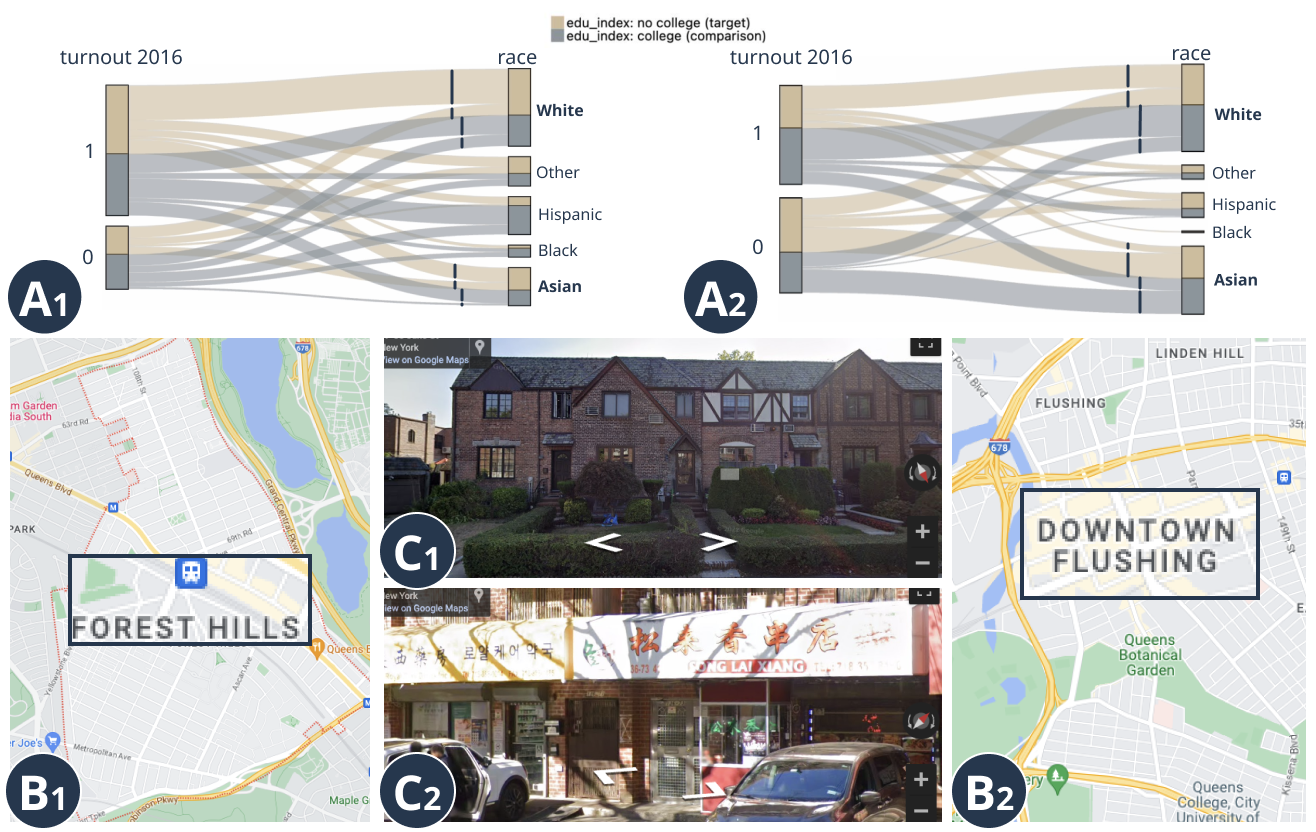} 
 \vspace{-0.5cm}
 \caption{
     The neighborhood effect was compared and verified in detail. 
     (A1) The \pcp revealed high voter turnout among Asians in cluster 2, especially Asians with college degrees.
     (A2) The \pcp indicated a low voting rate among Asians in cluster 5, regardless of their college attendance. 
     The maps and street views of cluster 2 (B1, C1) and cluster 5 (B2, C2) showed that they are located near Forest Hills and Flushing, respectively. 
 }
 \vspace{-0.3cm}
 \label{fig:case1-4}
\end{figure}

\textbf{Comparison and Verification.}
To delve deeper, $E_A$ compared social groups, selecting college graduates for comparison (\textbf{T5}). \pcp right bars confirmed that cluster 2 had an evenly distributed racial mix (Fig. \ref{fig:case1-4}-A1), whereas cluster 5 was predominantly White and Asian (Fig. \ref{fig:case1-4}-A2).
Strand widths showed higher voter turnout among White college graduates in both clusters, but Asian voter turnout was inconsistent. This emphasized the spatial heterogeneity of political landscapes and varying Asian political participation, leading $E_A$ to wonder about the exact locations of these clusters.
$E_A$ virtually toured clusters 2 and 4 via the \streetview (\textbf{T6}) and identified cluster 2 as Forest Hills (Fig. \ref{fig:case1-4}-B1, C1), a diverse, family-friendly neighborhood. 
He noted the northwest-southeast orientation of the boundary and inferred a strong localized spillover of voting. Cluster 5 overlapped with Flushing (Fig. \ref{fig:case1-4}-B2, C2), known for its Chinatown and high renter population, and suggested weaker social ties could limit voting spillover. 
Finally, he reflected, \textit{``An ethnic enclave leads to lower voter turnout among the Asian population, but a diverse community can boost political participation. 
I'll examine more U.S. regions for this pattern. ''}

\secondround{\textbf{Expert Feedback.}
We gathered feedback from our experts in Section~\ref{sec:02_Background_Requirement_Analysis} ($E_A$-$E_C$).
All experts thought \systemname provided a streamline for effectively exploring the neighborhood effect on the political engagement of minority social groups, offering valuable insights for future studies. 
Specifically, $E_B$ suggested that the analysis framework could be generalized to study the neighborhood effect of income inequality. 
$E_A$ appreciated the projection bars and spillover glyphs that make the model results intuitive to explore and understand.
$E_C$, as a GIS expert in econometrics, suggested including rigorous statistical testing in specific steps in the framework, yet she still appreciated the usefulness of \systemname to generate hypotheses for further studies.}
\section{Conclusion}
In this study, we present \systemname, an interactive visual analytics system to help social scientists model, explore, and verify neighborhood effects on different social groups. 
Future research using \systemname will include expanding options for statistical testing and visual encoding, incorporating diverse datasets for broader social science research, and enhancing the system's ability to handle time-varying data.

\balance
\end{spacing}

\acknowledgments{%
  This work was partially supported by RGC GRF grant 16210321.
}

\newpage
\bibliographystyle{abbrv-doi-hyperref}

\balance
\bibliography{template} 
\clearpage


\appendix 
\section{Appendix}

\subsection{Data Abstraction} \label{data_desc}

As neighborhood effect analysis over different social groups is a frequent research concern across multiple social science problems (\eg, partisan segregation \cite{brown2022partisan} and adoption of prosocial behavior ~\cite{baxter2022local}), we abstract the data format as follows to support versatile applications in different scenarios. 
Specifically, it consists of three types of data: subgroup, census, and meta datasets. 

\textit{Definition 1. (Subgroup Dataset):} A subgroup is a minimal group of individuals who share similar characteristics located in the same spatial unit (\eg, census block group (CBG) in the US ~\cite{cbgwiki}). A subgroup dataset $Sd$ is a collection of data tables for each year's subgroup data. 
Such a dataset is designed to maintain confidentiality at an individual level while supporting various applications. Each row $r$ contains the following information:
  \begin{compactitem}
    \item $r.population$: the number of individuals in the subgroup.
    \item $r.demographic$: demographic profiles (\eg, race and age group), usually treated as independent variables.
    \item $r.socioeconomic$: socioeconomic factors (\eg, employment status and household income), usually treated as another type of independent variable.
    \item $r.behavioral$: behavioral outcomes (\eg, political participation and financial behaviors) treated as dependent variables, recording the number of people who exhibited the behavior.
  \end{compactitem}

\textit{Definition 2. (Census Dataset):} a census dataset $Cd$ is a collection of separate data tables for each year's snapshot's aggregated data, where each row $r$ represents a spatial unit. The columns describe the census statistics (\eg, average household income and education attainment distribution) of the region and are treated as spatial independent variables of the environment.

\textit{Definition 3. (Meta Dataset):} a meta dataset $Md$ consists of shapefiles that include the geometry, coordinates, and other spatial information for each spatial unit in the individual and census datasets. These shapefiles enable the integration of spatial characteristics with the demographic, socioeconomic, and behavioral data present in the individual and census datasets. 

Together, these datasets ($Sd$, $Cd$, $Md$) provide a comprehensive and flexible data abstraction that allows for the exploration and analysis of various relationships between individuals.

\subsection{System Implementation}
\label{implementation}
The backend module is implemented in Python. It incorporates a data preprocessing algorithm (Fig. \ref{fig:framework}-A1), a Data Query Engine (Fig. \ref{fig:framework}-A2), and a streamlined data analysis pipeline (Fig. \ref{fig:framework}-A3 to A7). The data preprocessing algorithm transforms data collected from various sources into the specific data structure outlined in Section ~\ref{data_desc} and subsequently stores them in a MySQL database. The Data Query Engine aggregates summary statistics for subgroups based on the user-selected demographic attributes. The data analysis pipeline consists of a series of neighborhood effect analysis algorithms utilizing the geospatial analysis library, PySAL \cite{rey2009pysal}. Each submodule within the pipeline is encapsulated within a function, allowing for flexible interaction with the frontend via RESTful APIs.

The frontend module (Fig. \ref{fig:framework}-B1 to B6) is composed of six coordinated views. Utilizing technologies such as React.js ~\cite{react}, D3.js ~\cite{bostock2011d3}, and Plotly ~\cite{plotly}, it allows for real-time data visualization and supports interactive data analysis.

\subsection{Computational Complexity} \label{complexity}
The primary computational tasks in our pipeline include Geographically Weighted Regression (GWR), regionalization, and the calculation of spillover effects.
The GWR model commences with the selection of the optimal kernel bandwidth, followed by localized regression for each spatial unit. This process results in a computational complexity of $O(k^3n^2 \log n)$, where $k$ is the number of independent variables and $n$ is the number of spatial units \cite{li2020computational}.
The regionalization process integrates spatially constrained hierarchical clustering. The complexity of this operation can be estimated as $O(n^2)$ for constructing a distance matrix, and subsequently, $O(n^2 \log n)$ for the hierarchical clustering process.
The analysis of local spillover effects necessitates the calculation of a weighted sum of coefficients for each spatial unit, rendering a computational complexity of $O(kn)$, under the assumption that each spatial unit has an average of $k$ neighbors.
The system computes models and renders all visualization in real-time.
These computationally demanding tasks were integrated and evaluated on a computer equipped with an Intel Core i7 CPU and 16GB RAM. The algorithm effectively processed computations for 1,000 spatial units, each featuring 5 independent variables, in an average time of 20 seconds.
Our Python implementation additionally leverages parallel processing capabilities to optimize the performance of these computationally intensive operations.







\end{document}